\documentclass[twocolumn,10pt]{article} % here we use the article class, rather than elsarticle

\usepackage[square,numbers,sort&compress,comma]{natbib}

\usepackage{amsmath}
\usepackage{amssymb}
\usepackage{caption}
\usepackage{graphicx}
\usepackage{latexsym}
\usepackage{times,hyperref}
\usepackage[pagewise]{lineno}
\usepackage[version=4]{mhchem}
\topmargin - 12pt % might need to be set to 0pt for some installations
\renewenvironment{abstract}%
              {% - begin definition
               \small% - select font
               {\bfseries \abstractname}% - select font
               \par% - end a paragraph (skip \parsep)
               \vspace{10pt}% - add vertical space
              }% - complete definition

\renewcommand\abstractname{Abstract}

\newcommand{\nomenclature}% - name of command
              [1]% - number of arguments
              {% - begin definition
               \bgroup% - begin a local group
               \flushleft% - turn on flushleft option
               \small\bf% - select font
               #1% - insert title text
               \par% - end a paragraph (skip \parsep)
               \egroup% - terminate local group
              }% - complete definition

\renewcommand{\section}% - name of command
              [1]% - number of arguments
              {% - begin definition
               \bgroup% - begin a local group
               \flushleft% - turn on flushleft option
               \small\bf% - select font
               \refstepcounter{section}% - increment counter
               \arabic{section}. #1% - insert title text
               \par% - end a paragraph (skip \parsep)
               \egroup% - terminate local group
              }% - complete definition

\renewcommand{\subsection}% - name of command
              [1]% - number of arguments
              {% - begin definition
               \bgroup% - begin a local group
               \flushleft% - turn on flushleft option
               \small\em% - select font
               \refstepcounter{subsection}% - increment counter
               \arabic{section}.% - insert title text
               \arabic{subsection}. #1% - insert title text
               \par% - end a paragraph (skip \parsep)
               \egroup% - terminate local group
              }% - complete definition

\renewcommand{\subsubsection}% - name of command
              [1]% - number of arguments
              {% - begin definition
               \bgroup% - begin a local group
               \flushleft% - turn on flushleft option
               \small\em% - select font
               \refstepcounter{subsubsection}% - increment counter
               \arabic{section}.% - insert title text
               \arabic{subsection}.% - insert title text
               \arabic{subsubsection}. #1% - insert title text
               \par% - end a paragraph (skip \parsep)
               \egroup% - terminate local group
              }% - complete definition

  \newcommand{\acknowledgement}% - name of command
              [1]% - number of arguments
              {% - begin definition
               \bgroup% - begin a local group
               \flushleft% - turn on flushleft option
               \small\bf% - select font
               #1% - insert title text
               \par% - end a paragraph (skip \parsep)
               \egroup% - terminate local group
              }% - complete definition

  \newcommand{\sectionbib}% - name of command
              [1]% - number of arguments
              {% - begin definition
               \bgroup% - begin a local group
               \flushleft% - turn on flushleft option
               \small\bf% - select font
               #1% - insert title text
               \par% - end a paragraph (skip \parsep)
               \egroup% - terminate local group
              }% - complete definition

\begin{document}

\title{\LARGE GPU-accelerated Large Eddy Simulation of turbulent stratified flames with machine learning chemistry }

\author{{\large Min Zhang$^{a,b}$, Runze Mao$^{a,b}$, Han Li$^{a,b}$, Ruixin Yang$^{a,b}$,  Zhi X. Chen$^{a,b,*}$}\\[10pt]
        {\footnotesize \em $^a$College of Engineering, Peking University, Beijing, 100871, P.R. China}\\[-5pt]
        {\footnotesize \em $^b$AI for Science Institute, Beijing, 100080, P.R. China}\\[-5pt]
        }

\date{}

% -------------------------------------------------------------------- %
% -------------------------------------------------------------------- %
% -------------------------------------------------------------------- %

\small
\baselineskip 10pt

% -------------------------------------------------------------------- %
% -------------------------------------------------------------------- %
% -------------------------------------------------------------------- %

\twocolumn[\begin{@twocolumnfalse}
\vspace{50pt}
\maketitle
\vspace{40pt}
\rule{\textwidth}{0.5pt}
\begin{abstract} % 100 to 300 words.
Stratified premixed combustion, known for its capability to expand flammability limits and reduce overall-lean combustion instability, has been widely adopted to comply with increasingly stringent environmental regulations. Numerous numerical simulations with different combustion models and mesh resolutions have been conducted on laboratory-scale flames to further understand the stratified premixed combustion. However, the trade-off between the high-fidelity and low computational cost for simulating laboratory-scale flames still remains, particularly for those combustion models involving direct coupling of chemistry and flow. In the present study, a GPU-based solver is employed to solve partial differential equations and calculate the thermal and transport properties, while an artificial neural network~(ANN) is introduced to replace reaction rate calculation. Particular emphasis is placed on evaluating the proposed GPU-ANN approach through the large eddy simulation of the Cambridge stratified flame. The simulation results show good agreement for the flow and flame statistics between the GPU-ANN approach and the conventional CPU-based solver with direct integration~(DI). The comparison suggests that the GPU-ANN approach can achieve the same level of accuracy as the conventional CPU-DI solver. In addition, the overall speed-up factor for the GPU-ANN approach is over two orders of magnitude. This study lays the potential groundwork for fully resolved laboratory-scale flame simulations based on detailed chemistry with much more affordable computational cost.
\end{abstract}
\vspace{10pt}
\parbox{1.0\textwidth}{\footnotesize {\em Keywords:} Stratified flames; GPU; Machine learning; Finite rate chemistry; Large eddy simulation}
\rule{\textwidth}{0.5pt}
\vspace{10pt}

*Corresponding author.\\
\textit{E-mail address:} chenzhi@pku.edu.cn (Zhi X. Chen).

\end{@twocolumnfalse}] 

\clearpage

\clearpage

% \linenumbers

\section{Introduction\label{sec:introduction}} \addvspace{10pt}

In response to the growing stringency of emission regulations and standards, numerous combustion applications, such as gas turbines and industrial furnaces, operate in a very lean-premixed mode which is prone to result in combustion instability and extinction~\cite{Dunn:2011}.  Utilizing technology involving stratified premixed combustion has been shown to extend the flammability limits and lower the overall-lean combustion instability~\cite{Alkidas:2007, Masri:2015}. %Detailed reviews on stratified premixed combustion are available in~\cite{Masri:2015,Lipatnikov:2017}. 

Recently, the Cambridge stratified burner, designed by Sweeney et al.~\cite{Sweeney:2012, Sweeney:2012a}, has been widely studied to investigate the stratified flame due to its full set of experimental data available in terms of velocity, temperature, and species distribution. A large amount of simulations on the Cambridge stratified flames were reported with a focus on using different combustion models, such as artificial thickened flame~(ATF)~\cite{Inanc:2021}, flame generated mamifolds~(FGM)~\cite{Zhang:2021a}, transport probability density function~(tPDF)~\cite{Turkeri:2021}, and partially stirred reactor~(PaSR)~\cite{Qian:2022}. All the aforementioned combustion models have been shown to reasonably capture the distribution of species and predict the major combustion properties. However, Proch et al.~\cite{Proch:2017} argued that one notable challenge associated with using different combustion models arises from the intricate interplay between sub-filter closures for unresolved velocity scales and sub-filter flame wrinkling. This complexity introduces the potential for errors to either compensate or amplify. 
Therefore, it is challenging to make a clear and definitive assessment of the combustion model performance. To this end, Inanc et al.~\cite{Inanc:2022} and Proch et al.~\cite{Proch:2017} performed flame-resolved simulations to investigate the Cambridge stratified flame where the flame scales are resolved in direct numerical simulation~(DNS) sense based on tabulated chemistry with acceptable computational cost. They concluded that the major physical properties of the flame-resolved simulation are comparable to that of classical DNS of much simpler flame configurations. However, it should be noted that flame-resolved simulation is still subject to the flamelet model assumption. As an example, species with slower chemical reactions, such as acetylene and nitrogen dioxide, require more time to respond to changes in the dissipation rate thus resulting in an under-prediction of species concentration. It is apparent that there is still a trade-off between high fidelity and low computational cost for simulating laboratory-scale flames.   

To tackle this trade-off of high fidelity versus efficiency, the recent rapid development in artificial intelligence (AI), especially in machine learning (ML), has provided a new perspective for accelerating detailed finite-rate chemistry with high accuracy~\cite{Ihme:2022}. 
%Detailed reviews involving ML in the combustion field are available in~\cite{Ihme:2022}. 
A common approach to accelerate finite-rate chemistry is to replace the expensive direct integration of reaction rate with an artificial neural network~(ANN)~\cite{Christo:1996, Blasco:1998}. The essence of this approach lies in collecting training data that enable the ANN to mimic the chemical mechanism with the utmost accuracy. Therefore, the emphasis has been mainly placed on the investigation of using different sampling approaches, such as manifold sampling~\cite{Maas:1992}, Monte Carlo method, and multi-scale sampling~\cite{Zhang:2022b}. Recently, Readshaw and Ding et al.~\cite{Readshaw:2023, Ding:2021} proposed a hybrid flamelet/random data and multiple multilayer perceptrons~(HFRD-MMLP) method to simulate the Cambridge stratified flame. This method demonstrated an excellent accuracy against direct integration and the time spent on the reaction source term is reduced by a factor of fourteen. However, the overall speed-up factor is around four. It is hence not difficult to speculate that a majority of the central processing unit~(CPU) hours were spent on solving the species and flow transport equations, especially when eight Eulerian stochastic fields were employed in their study. The computational cost spent on those transport equations will become more prominent with respect to the increasing numbers of species and stochastic fields. 

Graphic processing units~(GPU) designed for the highly parallel process of graphics rendering can significantly accelerate computational fluid dynamics~(CFD) simulations~\cite{Prez:2018, Bielawski:2023}. 
Meanwhile, incorporating the ANN model into a GPU-based CFD solver will further alleviate the aforementioned bottleneck of computational cost while maintaining high fidelity. These simulation acceleration techniques have been widely explored in separate efforts and 
to the best knowledge of the authors, simulations conducted using an integrated GPU-ANN approach have rarely been reported, certainly not for laboratory-scale turbulent flames.
With this motivation, the objective of the present study is first to formulate an ANN model similar to that in~\cite{Readshaw:2023} but with the PaSR combustion model, and then couple it with a fully GPU-accelerated solver~\cite{Mao:2023,Mao:2023a} based on OpenFOAM to perform large eddy simulation~(LES) of the Cambridge stratified SWB5 flame. The accuracy and computational performance are thoroughly assessed against detailed experimental data and numerical results obtained using the high-precision CVODE chemistry integrator. 

The remainder of this paper is structured as follows: GPU acceleration methodology and framework are briefly introduced in Section 2. Training data and model generation are presented in Section 3. The evaluation of the integrated GPU-ANN approach in the Cambridge Stratified flame is performed in Section 4. Lastly, concluding remarks are provided in Section 5.

\section{GPU acceleration methodology and framework\label{sec:sections0}} \addvspace{10pt}

The computational code used here is part of the DeepFlame open-source framework~\cite{Mao:2023, Mao:2023a}, which has been recently developed to leverage GPU and ML techniques for combustion simulations. To avoid the CPU-GPU memory copy overhead, the entire computational process is executed on GPU, including machine learning chemistry, fully implicit pressure-based solving of Partial Differential Equations (PDEs) using the finite volume method (FVM), and computation of thermal and transport properties. The ML-related operations and the solving of linear systems are implemented based on the libTorch library and NVIDIA’s AmgX library, respectively. The other operations, such as FVM implicit discretisation and explicit computations, are realised through CUDA kernel functions, ensuring meticulous thread and memory management. To enable direct communication of the GPUs, the NVIDIA Collective Communication Library (NCCL) is adopted over the common MPI approach for multi-processor parallelism. In addition, various optimizations have been conducted to enhance computational performance and reduce GPU memory footprint. More implementation details can be found in~\cite{Mao:2023a}. 

\section{Training data and model generation\label{sec:sections}} \addvspace{10pt}

\subsection{Training data generation\label{subsec:subsection2}} \addvspace{10pt}

Readshaw and Ding et al.~\cite{Readshaw:2023, Ding:2021} proposed a sampling method based on HFRD. This method has been demonstrated to enable ANNs to mimic the chemical mechanism with the utmost accuracy. Following the basics of the HFRD, a brief training data generation procedure and its new developments are presented here. 

% The instantaneous reaction rates of a given mechanism, determined by local species concentration~($Y_{i}$), temperature~($T$), and pressure~($p$), can be solved by the following systems of ordinary differential equations (ODEs).
% \begin{equation}
% \Dot{\omega}_{k} = \frac{dy_{i}}{dt} = f(y_{i},T,p)
% \label{omega}
% \end{equation}
% Given a specific initial composition, the numerical integration of Eq.~\ref{omega} over a time step, $\delta t$, produces an input-output pair as follows:
% \begin{equation}
% y_{i}(t) = y_{i}(t + \delta t)
% \label{input}
% \end{equation}
The fundamental underpinning of replacing reaction rate calculation with an ANN is that the solution of the ordinary differential equations (ODEs) can be approximated using a non-linear optimization process, commonly referred to as training. 
The training dataset is created through sampling from flamelet simulation coupled with the generation of random data points. One-dimensional~(1-D) premixed flame simulations are first used to generate the initial data. The chemical mechanism employed here is the DRM19 which consists of 20 species and 85 reactions~\cite{Kazakov:2023}. 
With the consideration of flammability limits and equivalence ratio~($\phi$) range in the target simulation case~(SWB5), a total of 62 flamelet simulations within the $\phi$ range of (0.45, 1.10) are performed. The main simulation details on 1-D flame including mesh size, computational domain, and flame initialization, are similar to~\cite{Readshaw:2023}. However, the non-unity diffusivities here are taken into account by employing the Hirschfelder and Curtiss mixture-averaged transport model. 
The data is randomly sampled throughout the domain as the simulations progress from initialization to a steady state. Concurrently, a minimum temperature threshold of $500\,\mathrm{K}$ is set to prevent the collection of a large amount of non-reactive data. 

\begin{figure}[t]
\centering
\includegraphics[width=192pt]{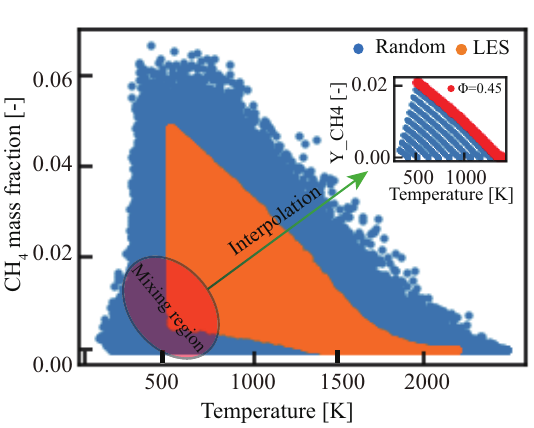}
\caption{Comparison of scatter plots between the random dataset and LES data from SWB5 in $T$-\ce{CH4} space.}
\label{intepolation}
\end{figure}

Following the 1D simulations, the flamelet dataset serves as the foundation for the generation of random data. For every composition within the flamelet dataset, a new random composition is generated through the following procedures:  
\begin{equation}
T' = T + \alpha c (T_{max} - T_{min}),
\label{Trand}
\end{equation}
\begin{equation}
p' = p + \beta c (p_{max} - p_{min}),
\label{Prand}
\end{equation}
\begin{equation}
y'_{N_{2}} = y_{N_{2}} + \alpha c (y_{N_{2},max} - y_{N_{2},min}),
\label{N2rand}
\end{equation}
\begin{equation}
y'_{j} = y_{j}^{(1+ \gamma c )},
\label{Yrand}
\end{equation}
where $c$ is random number within [-1,1] with uniform distribution. $\alpha$, $\beta$, and $\gamma$ are set to 0.125, 10.0, and 0.1, respectively. 
While the generated random data after the above procedures roughly covers the target area of composition space, it fails to account for the mixing between extremely lean reactants and high-temperature products, primarily due to the entrainment resulting from large eddy structures. To address this issue, the linear interpolation between the equilibrium compositions of the lowest $\phi = 0.45$ and the compositions of air at ambient conditions is employed for all species. In addition, The randomly generated data based on the aforementioned method must conform to multiple specified constraints. Initially, the mass fractions of species must sum to unity. This condition is guaranteed by normalizing the species mass fraction except for nitrogen~(\ce{N2}) to $1-y'_{N_{2}}$. Furthermore, the molar element ratio and equivalence ratio of each composition state in the random dataset must have appropriate values. It should be noted that the non-unity Lewis number is employed in the present study, the range of these values is hence set to differ from those used in ~\cite{Readshaw:2023}. The detailed constraints are given in Table~\ref{table:consrtraint}. 
As an illustration, Figure~\ref{intepolation} shows the $T$-\ce{CH4} space spanned by the obtained random data after the above procedures and LES simulation data of the SWB5. In the upper right corner of the figure, the scheme of linear interpolation between the lowest $\phi = 0.45$ at equilibrium state and air at ambient condition is presented.
Once the random dataset is generated, every composition within the random dataset will be integrated using the CVODE integrator~\cite{Brown:1989} over a time step of $1\,\mathrm{\mu s}$ to get the target output. In the present study, a total number of 518000 input-out pairs are generated as the final training dataset.  

\begin{table}[t]\footnotesize
\caption{Constraints for random data generation.}
\centering
\begin{tabular}{ l c c c}
%\specialrule{.1em}{.1em}{.1em}
\hline
 & \begin{tabular}[c]{@{}c@{}} \text{H/C} ratio\end{tabular} 
  &\begin{tabular}[c]{@{}c@{}} \text{O/N} ratio\end{tabular}
 &\begin{tabular}[c]{@{}c@{}} \text{$\phi$}\end{tabular}   \\ 
%\multirow{2}{*}{Multirow}&X\\
%\specialrule{.05em}{.1em}{.1em}
\hline
Minimum  & 2.65 &0.254    &0.20\\
Maximum  &4.67  &0.32   & 1.15\\
%\specialrule{.1em}{.1em}{.1em}
\hline
\end{tabular}
\label{table:consrtraint}
\end{table}

\subsection{ANN Model\label{sec:figtabeqn}} \addvspace{10pt}

% Since the different species exhibit disparate orders of magnitude in thermochemical phase space throughout the evolution of the chemical system, the Box-Cox transformation (BCT), proposed by Box and Cox (1964)~\cite{Box:1964}, is introduced here to represent multi-scale data by O(1) quantity and avoid the singularity arising from the log transformation when the data approaches zero. 
% \begin{equation}
% f(x) =  \begin{cases}
%             \frac{x^{\lambda -1}}{\lambda} \\
%       \text{log}(x) 
%          \end{cases}
% \label{bct}
% \end{equation}
% where $\lambda = 0.1$ is adopted in the present study. The mass fraction of the chemical species within [0,1] is mapped to [$-1/ \lambda$, 0] after the BCT.

\begin{figure}[t]
\centering
\includegraphics[width=192pt]{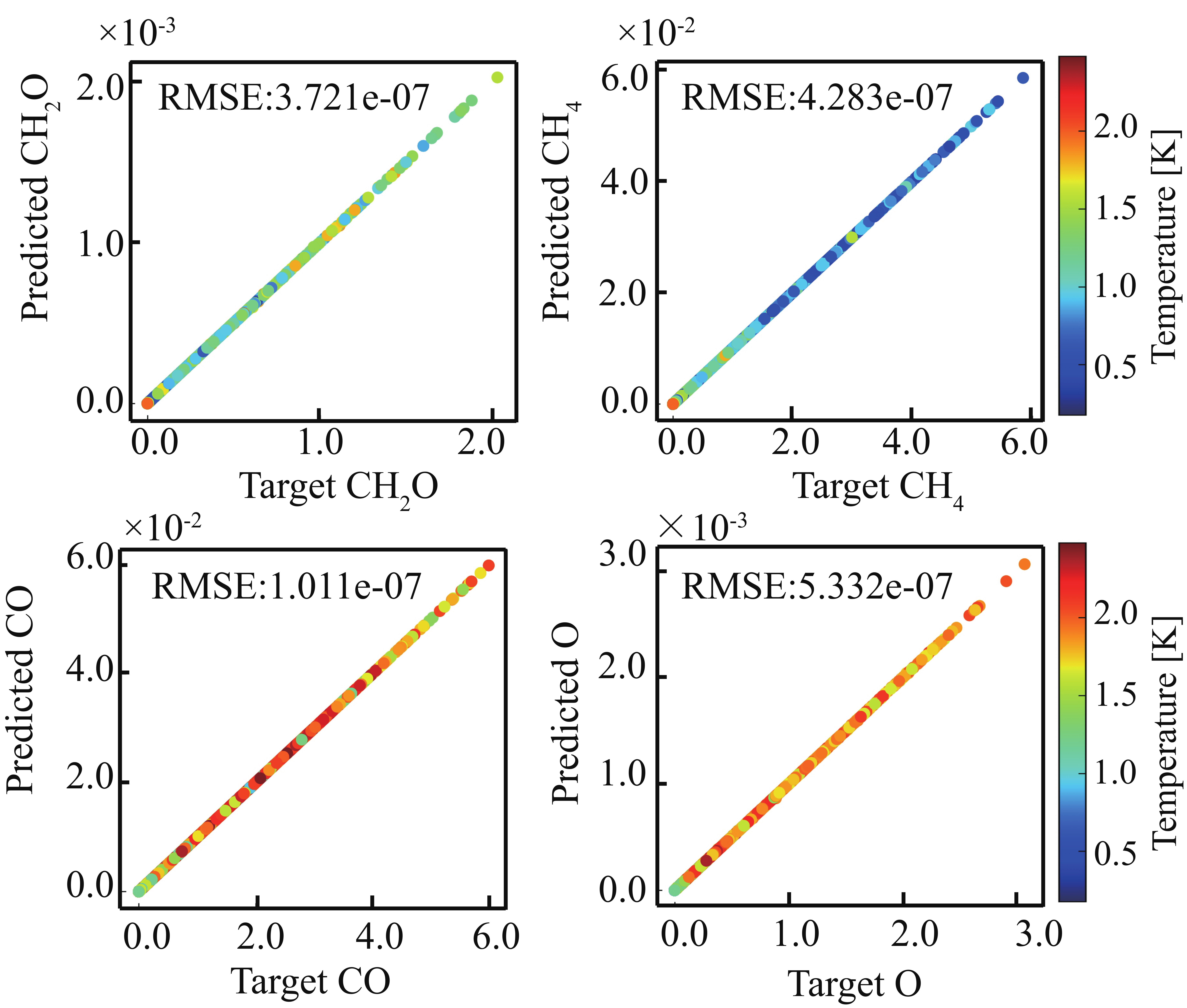}
\caption{MLP predictions for \ce{CH2O}, \ce{CH4}, \ce{CO}, and \ce{O}. Each dot is colored by the temperature of the sample.}
\label{MLP}
\end{figure}

Considering that the instantaneous reaction rates are determined by local species mass fractions~($Y_{i}$), temperature~($T$), and pressure~($p$), the ANN input layers are hence represented as $\boldsymbol{x}(t)=\{T(t), p(t), \mathcal{F}(\boldsymbol{Y}(t))\}$.
The output layers comprise the change in species mass fraction over a given time step, represented as $\boldsymbol{u}^*(t)=\boldsymbol{Y}(t+\Delta t) - \boldsymbol{Y}(t)$. 
The Box-Cox transformation $\mathcal{F}(x)$~\cite{Box:1964}, is introduced here to represent multi-scale species mass fraction by $\mathcal{O}(1)$ quantity and avoid the singularity arising from the log transformation when the data approaches zero.
In addition, the training dataset with the sample size of $N$, denoted as $D=\{\boldsymbol{x_i},\boldsymbol{u^*_i}\}^N_{i=1}$, undergoes $Z$-score normalization. Individual training and prediction are conducted for each species mass fraction, ensuring a high accuracy. Each ANN comprises hidden layers with 1600, 800, and 400 perceptrons.
The Gaussian Error Linear Unit (GELU) is employed as an activation function for the network, while the hyperparameter optimization is conducted through the Adam algorithm.

A commonly used loss function to constrain the output of the ANN is expressed as follows:
\begin{equation}
\begin{split}
    \mathcal{L}&=\frac{1}{N} \sum_{i=1}^N \left \vert \boldsymbol{u}^*_i - \boldsymbol{u}_i \right \vert
\end{split}
\label{Loss}
\end{equation}
where ${u}_i$ is the ANN output. In addition, three principles in terms of mass fraction unity conservation, energy conservation, and considerations related to heat release rate, are incorporated into the loss function to achieve a higher precision. The prior assessment of the prediction performance for this trained ANN is carried out based on another random dataset generated by repeating all the procedures outlined in Section 3.1. A good agreement between the predicted and target values is achieved, as shown in Fig.~\ref{MLP}. Therefore, this trained ANN is applicable for a posteriori assessment in the Cambridge Stratified flame.

\section{Application to Cambridge Stratified flame\label{sec:extext}} \addvspace{10pt}

\subsection{Case description and numerical methods} \addvspace{10pt}

The Cambridge burner is characterized by a central bluff-body enveloped by two co-annular premixed methane-air mixture streams, accompanied by a co-flow of air maintained under ambient conditions. As the objective of the present study is to assess the performance of the integrated GPU-ANN approach in a laboratory-scale flame, only the moderately stratified case SWB5 is studied here. More details on the experiment can be found in~\cite{Sweeney:2012, Sweeney:2012a}. For brevity, the main initial parameters and boundary conditions for the SWB5 case are listed in Table~\ref{table:swb5}.

\begin{table}[t]\footnotesize
\caption{Main parameters for Cambridge burner SWB5.}
\centering
\begin{tabular}{ l c c c c c c c}
%\specialrule{.1em}{.1em}{.1em}
\hline
\begin{tabular}[c]{@{}c@{}} \text{$U_{o}$} \\ \text{$\mathrm{m/s}$}\end{tabular}  & \begin{tabular}[c]{@{}c@{}} \text{$U_{i}$} \\ \text{$\mathrm{m/s}$}\end{tabular}  & \begin{tabular}[c]{@{}c@{}} \text{$U_{co}$} \\ \text{$\mathrm{m/s}$}\end{tabular}  & \begin{tabular}[c]{@{}c@{}} \text{$\phi_{i}$} \\- \end{tabular}     & \begin{tabular}[c]{@{}c@{}} \text{$\phi_{o}$} \\ - \end{tabular} & \begin{tabular}[c]{@{}c@{}} \text{$\phi_{Co}$} \\ - \end{tabular} & \begin{tabular}[c]{@{}c@{}} \text{Swirl} \\ \% \end{tabular} \\
\hline
8.3  & 18.7    & 0.4   &  1.0  & 0.5 &0.0 &0.0 \\
%\specialrule{.1em}{.1em}{.1em}
\hline
\end{tabular}
\label{table:swb5}
\end{table}

\begin{figure}[t]
\centering
\includegraphics[width=192pt]{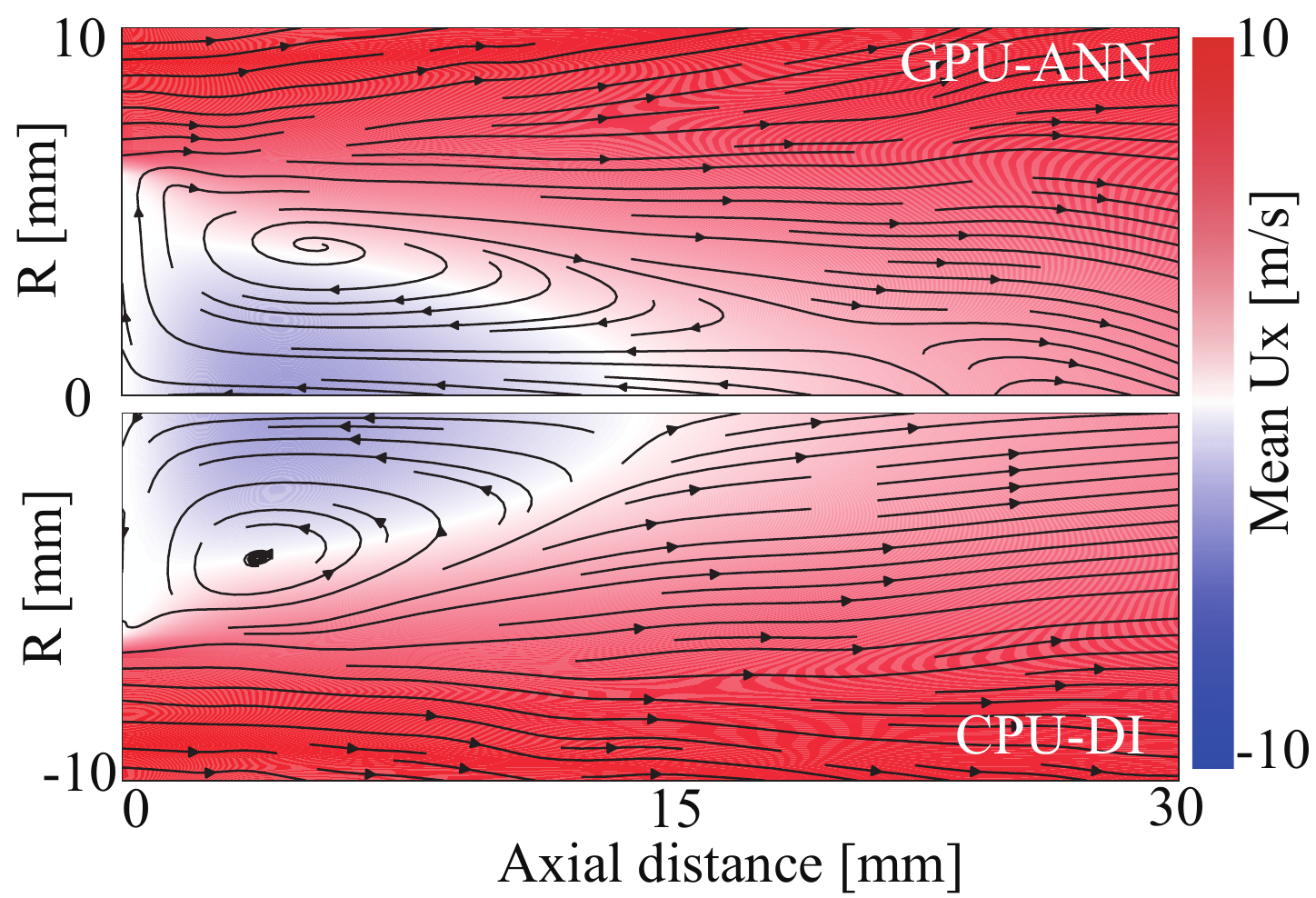}
\caption{Mean axial velocity field and streamlines for the CPU-DI and GPU-ANN cases.}
    \label{Umean}
\end{figure}

\begin{figure*}[ht!]
\centering
\vspace{-0.4 in}
\includegraphics[width=130mm]{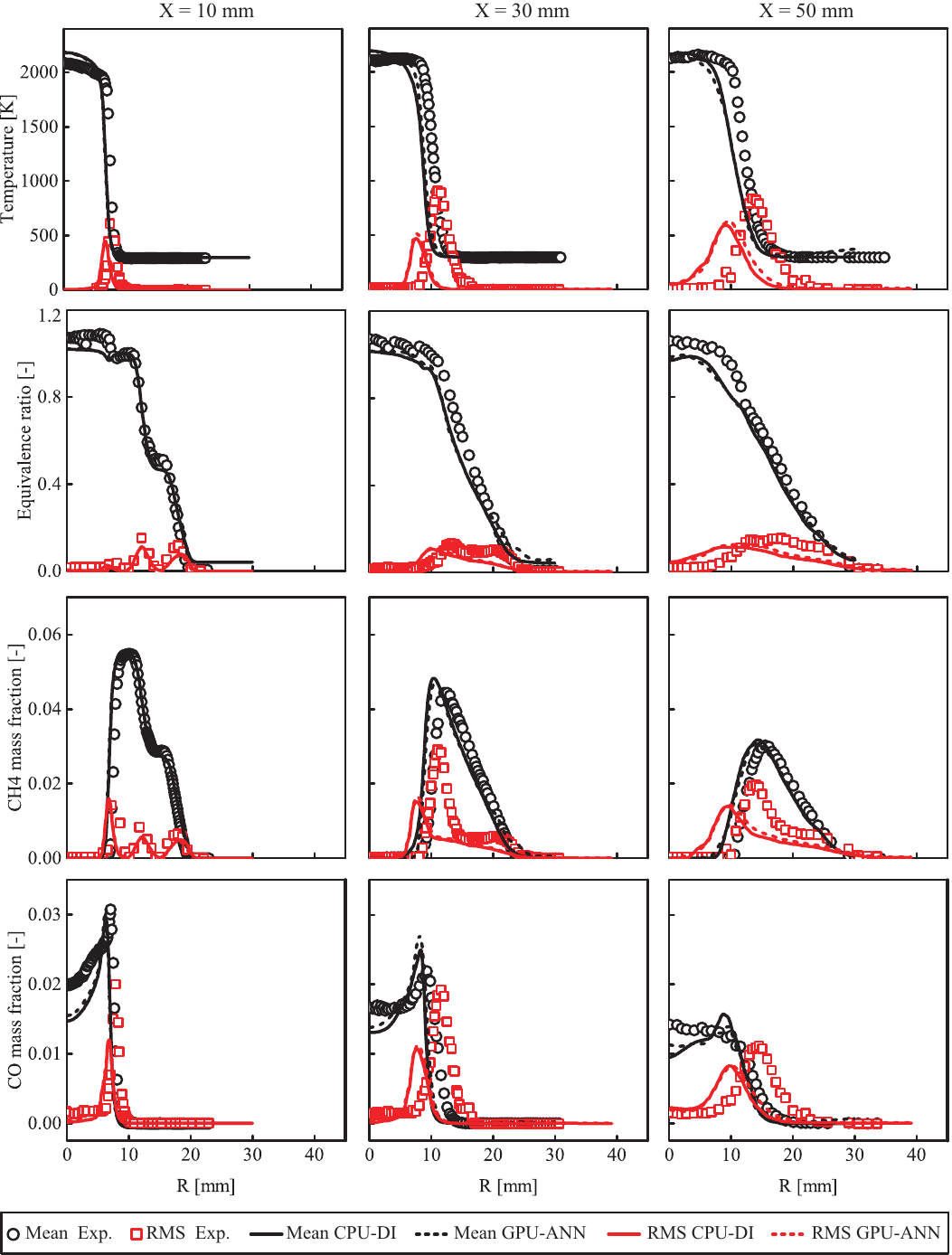}
\vspace{10 pt}
\caption{Radial profiles of mean and RMS for temperature, equivalence ratio, \ce{CH4}, and \ce{CO} mass fraction at three different axial locations $\text{X} = 10\,\mathrm{mm}, 30\,\mathrm{mm}, 50\,\mathrm{mm}$ in the experiment, CPU-DI, and GPU-ANN cases.}
\label{major}
\end{figure*}

\begin{figure*}[t]
\centering
\vspace{-0.4 in}
\includegraphics[width=130mm]{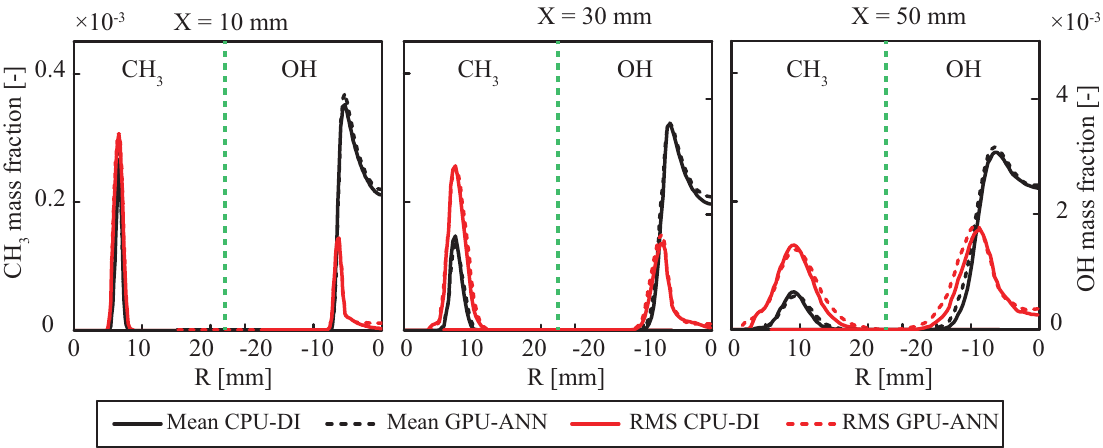}
\vspace{10 pt}
\caption{Radial profiles of mean and RMS for minor species \ce{CH3}, and \ce{OH} mass fraction at three different axial locations~$X = 10\,\mathrm{mm}, 30\,\mathrm{mm}, 50\,\mathrm{mm}$ in the CPU-DI and GPU-ANN cases.}
\label{minor}
\end{figure*}

The computational domain is a cylinder with a length of $300\,\mathrm{mm}$ and $100\,\mathrm{mm}$ in the axial and radial directions. A nonuniform grid with a total of 2.5 million cells is employed. The mesh refinement is specifically applied to the axial direction near the jet exit and to the radial direction within the shear layer situated between the fuel jets and the coflow. The average mesh spacing in the refined region is around $0.25\,\mathrm{mm}$, which is sufficient to create a high resolution in the flame region, according to the grid size used in~\cite{Qian:2022, Zhang:2021a}. In this study, the gas-phase flow field described in the Eulerian framework is resolved using the Favre-filtered compressible Navier-Stokes equations, coupled with a Smagorinsky sub-grid-scale model. The turbulence-chemistry-interaction is modelled using the PaSR combustion model wherein the characteristic chemical time is calculated as the slowest formation rate of species and the mixing time scale is calculated as the geometric mean of the Taylor scale and the Kolmogorov scale.  
The spatial discretization for momentum lies in a second-order scheme. No-slip boundaries are employed on all walls. Pseudo-turbulent fluctuations generated using a synthetic eddy turbulence generator~\cite{Kornev:2007} are imposed at the fuel inlet. The integral length scale and fluctuation level are the same as those used in~\cite{Proch:2017}. The mean and root mean square~(RMS) statistics are time-averaged over roughly two flow-through times with respect to the inner jet velocity through the axial distance of $50\,\mathrm{mm}$. Meanwhile, an additional averaging is performed in the azimuthal direction.

\subsection{Results and discussion} \addvspace{10pt}

In the present study, two simulation cases for the SWB5 are carried out with two different approaches. In one case, the flow and scalar transport equations are solved in a CPU-based solver and the reaction source terms are integrated using the CVODE solver. In another case, the corresponding equations are solved in a GPU-based solver, and a trained ANN model is employed to predict the reaction rates. To facilitate the description, the abbreviations CPU-DI and GPU-ANN are introduced here for these two cases. 

Figure~\ref{Umean} illustrates the mean axial velocity fields and their streamlines for both the CPU-DI and GPU-ANN results. One main feature of the flow structure is the large recirculation region induced by the bluff body which contributes to the stabilization of the flame. Both CPU-DI and GPU-ANN results show the recirculation region extends around $15\,\mathrm{mm}$ downstream of the fuel exit of the burner. The large vertices indicated by streamlines are almost identical for both the CPU-DI and GPU-ANN simulation results. 

While Fig.~\ref{Umean} has already shown that the flow structure can be accurately captured with GPU-ANN, it is unclear whether the stratified flame structure can be accurately simulated in the GPU-ANN case. This will be explored in the subsequent discussion. Figs.~\ref{major} and~\ref{minor} show the mean and RMS statistics for these two simulation cases. The mean and RMS values are indicated by black and red colors, respectively. The solid lines and dashed lines denoted the results obtained from CPU-DI and GPU-ANN cases, while the circles and squares denote the experimental mean and RMS data~\cite{Sweeney:2012, Sweeney:2012a}, respectively. In Fig.~\ref{major}, a comparison is presented among the experimental data, CPU-DI, and GPU-ANN simulation results regarding the radial distribution of temperature, equivalence ratio, as well as major species methane~(\ce{CH4}) and carbon monoxide (\ce{CO}). It is evident that the simulation results are almost identical between CPU-DI and GPU-ANN, indicating that the GPU-ANN approach can reproduce the major combustion properties simulated by the conventional CPU-DI solver. When comparing the simulation results to the experimental data, the simulated results show a good agreement with the experimental data at $\text{X} = 10\,\mathrm{mm}$, with only a slight difference observed in the recirculation region above the bluff body. The possible reason is likely due to the adiabatic boundary employed in the present study. This is supported by the simulation results from Mercier et al.~\cite{Mercier:2015} whose simulations showed that the adiabatic boundary overestimates the temperature by about $150\,\mathrm{K}$ compared to the non-adiabatic boundary. At $ \text{X} = 30\,\mathrm{mm}, 50\,\mathrm{mm}$, one can see that the temperature profiles are narrower than that in the experiment, indicating a slower flame spread in the simulation. This consequently results in a shift of the peak position for the RMS. The reason can be attributed to two factors. First, the turbulence inlet conditions adopted from the simulation~\cite{Proch:2017} should be further tuned to satisfy the current simulation cases. This is reflected by the lower prediction of RMS peak values. Second, the combustion reaction in the downstream region is overly suppressed by the PaSR combustion model. However, it should be noted that comparing the simulation results with the experimental data is not the main aim of this work. A further improvement to match the experimental data is out of the scope of the present study. Hence, the subsequent discussion will be focused on the comparison between the results in the GPU-ANN and CPU-DI cases.   

\begin{figure*}[t]
\centering
\vspace{-0.4 in}
\includegraphics[width=135mm]{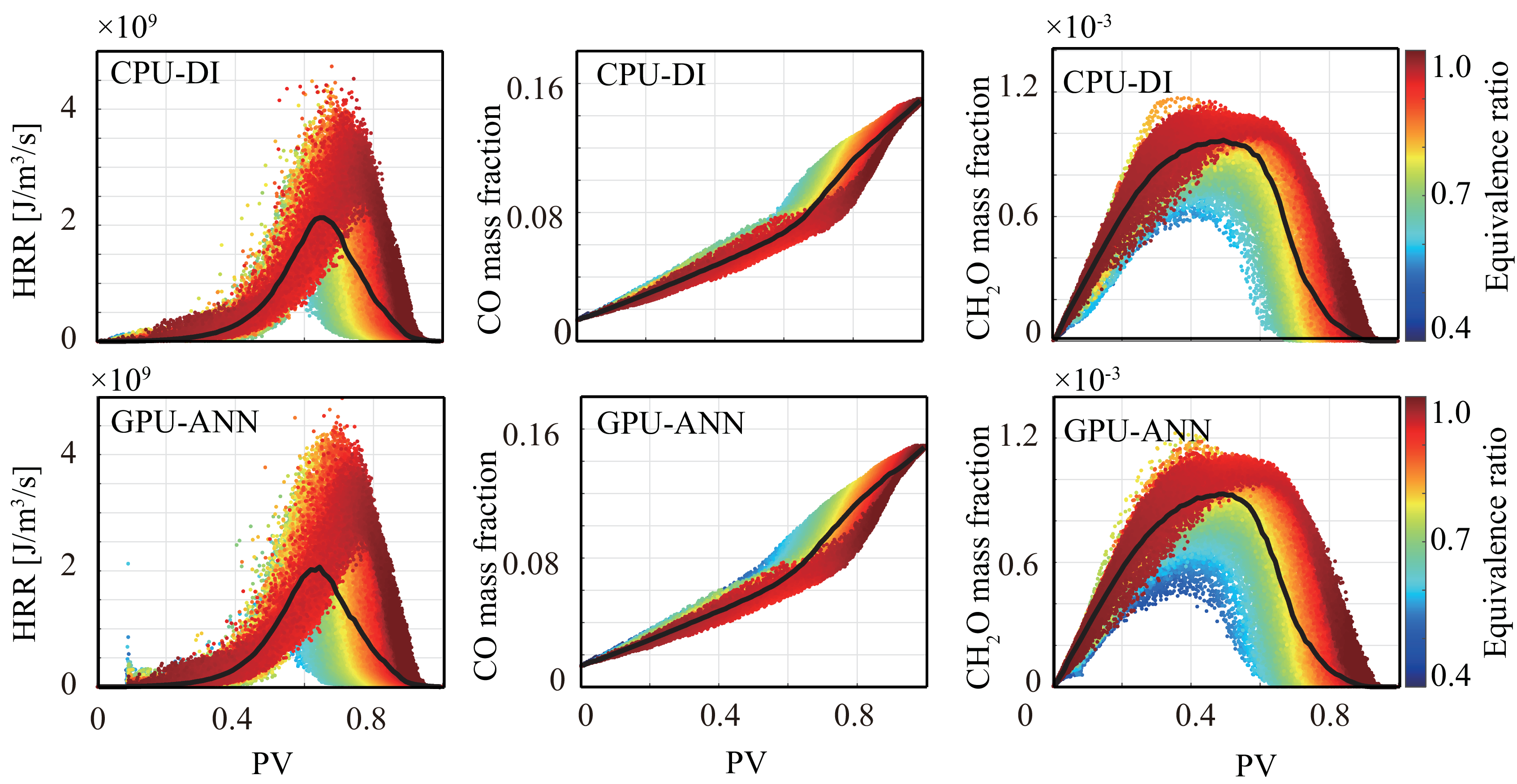}
\vspace{10 pt}
\caption{Instantaneous scatter plots of HRR, \ce{CO} and \ce{CH2O} mass fraction over progress variable for CPU-DI and GPU-ANN.}
\label{scatter}
\end{figure*}

In the chemical reaction system, the minor species span a broader range of orders of magnitude in thermochemical phase space, thus causing a greater challenge for the ANN prediction as compared to the major species. It is essential to further evaluate the accuracy of the GPU-ANN approach regarding the prediction of minor species. As illustrated in Fig.~\ref{minor}, the radial profiles of the minor species methyl group (\ce{CH3}) and hydroxide (\ce{OH}) in the CPU-DI and GPU-ANN cases at different axial locations are compared. It can be seen that the radial profiles for the CPU-DI and GPU-ANN are again in good agreement at all three axial locations. Some small differences can be observed near the mixing region where a strong interaction occurs between the fresh gas and hot combustion products due to the intense turbulent fluctuations. Nevertheless, these minor differences are deemed acceptable, and the overall profiles from GPU-ANN closely align with the results in the CPU-DI case. 
  
% \begin{figure}[t]
% \centering
% \includegraphics[width=192pt]{fig/Qdot_Umean.pdf}
% \caption{MLP predictions for \ce{CH2O}, \ce{CH4}, \ce{CO}, and \ce{O}. Each dot is colored by the temperature of the sample.}
% \label{contour}
% \end{figure}

The results in Figs.~\ref{major} and \ref{minor} have demonstrated that the flame structure for both the CPU-DI and GPU-ANN is identical only concerning the statistical mean and RMS. It is essential to assess the flame structure through single-shot data. Figure~\ref{scatter} compares the instantaneous scatters of the heat release rate~(HRR), \ce{CO} and formaldehyde (\ce{CH2O}) mass fraction colored by equivalence ratio over the progress variable~(PV) space between the CPU-DI and GPU-ANN cases. Their conditionally averaged values represented by black lines are also superimposed. It is seen that the scatter plots of HRR and \ce{CO}, along with their conditional mean values in the GPU-ANN case, demonstrate good agreement with those in the CPU-DI case. However, some discrepancies are observed for the scatter plots of \ce{CH2O}. One can see that more \ce{CH2O} with a low equivalence ratio in the GPU-ANN case are generated at the moderate combustion stage where the PV is around 0.4. Since \ce{CH2O} plays an important role in the preheat zone region, a further improvement in the accuracy of \ce{CH2O} may be crucial, particularly for flames with a high occurrence of local extinction and re-ignition.    

\begin{figure}[t]
\centering
\includegraphics[width=192pt]{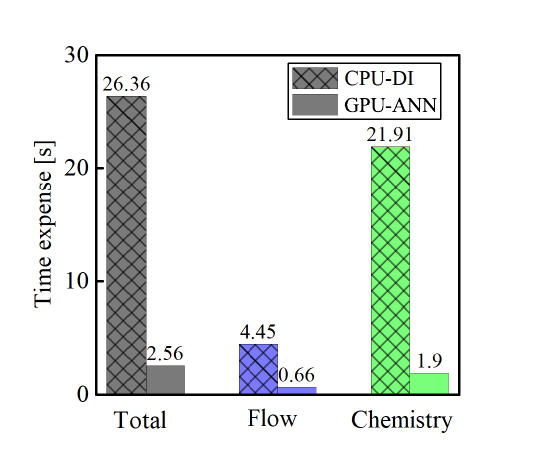}
\caption{Time expense for CPU-DI (32 cores) and GPU-ANN (1 card) cases.}
\label{timeCost}
\end{figure}

As described in the Introduction, the GPU-ANN approach aims to simulate the laboratory-scale flame with high-fidelity chemistry but at low computational cost.  
The accuracy of the GPU-ANN has already been evaluated in Figs.~\ref{Umean} to~\ref{scatter}, showing the same level of accuracy as the conventional CPU-DI solver. As for the computational cost, the total, flow, and chemistry time expenses of a single time step for the CPU-DI and GPU-ANN cases are compared in Fig.~\ref{timeCost}. The GPU-ANN case is carried out with one GPU card, while the CPU-DI case is performed with 32 CPU cores. Compared with CPU-DI, the time spent using GPU-ANN on the chemistry reaction and flow is reduced by 91.3\,\% and 85.2\,\%, respectively. This corresponds to a 1GPU-to-1CPU speed-up ratio of 369 and 216, respectively. The overall computation of a single step performed using GPU-ANN (1 card) is over 10 times faster than using CPU-DI (32 cores). 
This significant computational acceleration suggests that with the proposed GPU-ANN approach, high-fidelity LES/DNS simulation with finite-rate chemistry can be readily simulated with affordable resources that are available to a wider community. 
%When considering one GPU card versus one CPU core, the speed-up factor is approximately 416. 
%This suggests that there is no longer a bottleneck in resolving laboratory-scale flames in the DNS sense based on finite-rate chemistry with low computational cost.       

\section{Conclusions} \addvspace{10pt}

In the present study, a GPU-accelerated solver integrated with machine learning chemistry is comprehensively evaluated in a laboratory-scale flame. Specifically, an ANN model is trained based on 1D laminar flame solutions augmented with random perturbations. this model is then incorporated into a fully GPU-accelerated solver based on OpenFOAM to simulate the Cambridge stratified flame using large eddy simulation coupled with the PasR combustion model. 
The simulation results demonstrated that the flow and flame structures in terms of the statistical mean and RMS values and instantaneous scatters, are in good agreement between the GPU-ANN approach and the one using a conventional CPU solver with direct chemistry integration. 
%In the GPU-ANN, a GPU-based solver solves these equations and incorporates a trained MLP model for predicting reaction rates. The MLP models for each species mass fraction are trained with the random dataset which is generated using the sampling method proposed by Readshaw and Ding et al.~\cite{Readshaw:2023, Ding:2021}.
Furthermore, the computational acceleration by GPU-ANN is evaluated and the results show that the time spent on chemistry and flow calculation is reduced by 91.3\,\% and 85.2\,\%, respectively, when comparing one GPU card against 32 CPU cores. This corresponds to an overall speed-up factor of over two orders of magnitude.
This study showcases the potential of GPU-ANN approach and paves the way to solving laboratory-scale flames in the DNS sense based on finite-rate chemistry with acceptable computational cost.

%\section{Nomenclature and appendix\label{sec:nomapp}} \addvspace{10pt}

\acknowledgement{Declaration of competing interest} \addvspace{10pt}
The authors declare that they have no known competing financial interests or personal relationships that could have appeared to influence the work reported in this paper.

% \acknowledgement{Acknowledgments} \addvspace{10pt}

% The computation was performed using the High-Performance Computing Platform of CAPT of Peking University~(PKU).

%\acknowledgement{Supplementary material} \addvspace{10pt}

%No supplementary material.

% -------------------------------------------------------------------- %
% -------------------------------------------------------------------- %
% -------------------------------------------------------------------- %

 \footnotesize
 \baselineskip 9pt

% -------------------------------------------------------------------- %
% -------------------------------------------------------------------- %
% -------------------------------------------------------------------- %

\bibliographystyle{pci}
\bibliography{article}

\begin{thebibliography}{10}
\expandafter\ifx\csname url\endcsname\relax
  \def\url#1{\texttt{#1}}\fi
\expandafter\ifx\csname urlprefix\endcsname\relax\def\urlprefix{URL }\fi
\expandafter\ifx\csname href\endcsname\relax
  \def\href#1#2{#2} \def\path#1{#1}\fi

\bibitem{Dunn:2011}
D.~Dunn-Rankin, Lean combustion: technology and control, Academic Press (2011).

\bibitem{Alkidas:2007}
A.~C. Alkidas, Combustion advancements in gasoline engines, Ecol.\ Model. 48~(11) (2007) 2751--2761.

\bibitem{Masri:2015}
A.~Masri, Partial premixing and stratification in turbulent flames, Proc.\ Comb.\ Inst. 35~(2) (2015) 1115--1136.

\bibitem{Sweeney:2012}
M.~S. Sweeney, S.~Hochgreb, M.~J. Dunn, R.~S. Barlow, The structure of turbulent stratified and premixed methane/air flames {I}: {N}on-swirling flows, Combust.\ Flame 159~(9) (2012) 2896--2911.

\bibitem{Sweeney:2012a}
M.~S. Sweeney, S.~Hochgreb, M.~J. Dunn, R.~S. Barlow, The structure of turbulent stratified and premixed methane/air flames {II}: {S}wirling flows, Combust.\ Flame 159~(9) (2012) 2912--2929.

\bibitem{Inanc:2021}
E.~Inanc, A.~M. Kempf, N.~Chakraborty, Effect of sub-grid wrinkling factor modelling on the large eddy simulation of turbulent stratified combustion, Combust.\ Theo. and Model. 25~(5) (2021) 911--939.

\bibitem{Zhang:2021a}
W.~Zhang, S.~Karaca, J.~Wang, Z.~Huang, J.~van Oijen, Large eddy simulation of the {C}ambridge/{S}andia stratified flame with flamelet-generated manifolds: Effects of non-unity {L}ewis numbers and stretch, Combust.\ Flame 227 (2021) 106--119.

\bibitem{Turkeri:2021}
H.~Turkeri, X.~Zhao, M.~Muradoglu, Large eddy simulation/probability density function modeling of turbulent swirling stratified flame series, Phys.\ Fluids 33~(2) (2021) 025117.

\bibitem{Qian:2022}
X.~Qian, C.~Zou, H.~Lu, H.~Yao, Large-eddy simulation of {C}ambridge-{S}andia stratified flames under high swirl, Combust.\ Flame 244 (2022) 112241.

\bibitem{Proch:2017}
F.~Proch, P.~Domingo, L.~Vervisch, A.~M. Kempf, Flame resolved simulation of a turbulent premixed bluff-body burner experiment. part {I}: Analysis of the reaction zone dynamics with tabulated chemistry, Combust.\ Flame 180 (2017) 321--339.

\bibitem{Inanc:2022}
E.~Inanc, A.~Kempf, N.~Chakraborty, Scalar gradient and flame propagation statistics of a flame-resolved laboratory-scale turbulent stratified burner simulation, Combust.\ Flame 238 (2022) 111917.

\bibitem{Ihme:2022}
M.~Ihme, W.~T. Chung, A.~A. Mishra, Combustion machine learning: Principles, progress and prospects, Prog.\ Energy Combust.\ Sci. 91 (2022) 101010.

\bibitem{Christo:1996}
F.~C. Christo, A.~R. Masri, E.~M. Nebot, S.~B. Pope, An integrated {PDF}/neural network approach for simulating turbulent reacting systems, Proc.\ Comb.\ Inst. 26~(1) (1996) 43--48.

\bibitem{Blasco:1998}
J.~Blasco, N.~Fueyo, C.~Dopazo, J.~Ballester, Modelling the temporal evolution of a reduced combustion chemical system with an artificial neural network, Combust.\ Flame 113 (1998) 38--52.

\bibitem{Maas:1992}
U.~Maas, S.~B. Pope, Simplifying chemical kinetics: intrinsic low-dimensional manifolds in composition space, Combust.\ Flame 88 (1992) 239--264.

\bibitem{Zhang:2022b}
T.~Zhang, Y.~Yi, Y.~Xu, Z.~X. Chen, Y.~Zhang, E.~Weinan, Z.-Q.~J. Xu, A multi-scale sampling method for accurate and robust deep neural network to predict combustion chemical kinetics, Combust.\ Flame 245 (2022) 112319.

\bibitem{Readshaw:2023}
T.~Readshaw, L.~L. Franke, W.~Jones, S.~Rigopoulos, Simulation of turbulent premixed flames with machine learning-tabulated thermochemistry, Combust.\ Flame 258 (2023) 113058.

\bibitem{Ding:2021}
T.~Ding, T.~Readshaw, S.~Rigopoulos, W.~Jones, Machine learning tabulation of thermochemistry in turbulent combustion: {A}n approach based on hybrid flamelet/random data and multiple multilayer perceptron, Combust.\ Flame 231 (2021) 111493.

\bibitem{Prez:2018}
F.~E.~H. P{\'e}rez, N.~Mukhadiyev, X.~Xu, A.~Sow, B.~J. Lee, R.~Sankaran, H.~G. Im, Direct numerical simulations of reacting flows with detailed chemistry using many-core/{GPU} acceleration, Computers \& Fluids 173 (2018) 73--79.

\bibitem{Bielawski:2023}
R.~Bielawski, S.~Barwey, S.~Prakash, V.~Raman, Highly-scalable {GPU}-accelerated compressible reacting flow solver for modeling high-speed flowsn, Computers \& Fluids 235 (2023) 105972.

\bibitem{Mao:2023}
R.~Mao, M.~Lin, Y.~Zhang, T.~Zhang, Z.-Q.~J. Xu, Z.~X. Chen, Deepflame: A deep learning empowered open-source platform for reacting flow simulations, Comp.\ Phys.\ Commun. 291 (2023) 108842.

\bibitem{Mao:2023a}
R.~Mao, Y.~Wang, M.~Zhang, H.~Li, J.~Xu, X.~Dong, Y.~Zhang, Z.~X. Chen, An integrated framework for accelerating reactive flow simulation using {GPU} and machine learning models, arXiv preprint (2023).

\bibitem{Kazakov:2023}
A.~Kazakov, M.~Frenklach, Reduced reaction sets based on {GRI-MECH} 1.2, http://www.me.berkeley.edu/drm/.

\bibitem{Brown:1989}
P.~N. Brown, G.~D. Byrne, A.~C. Hindmarsh, Vode: A variable-coefficient ode solver, SIAM J.\ Sci.\ Stat.\ Comput. 10~(5) (1989) 1038--1051.

\bibitem{Box:1964}
G.~E. Box, D.~R. Cox, An analysis of transformations, J.\ Roy.\ Stat.\ Soc.\ B 26~(2) (1989) 211--243.

\bibitem{Kornev:2007}
N.~Kornev, E.~Hassel, Method of random spots for generation of synthetic inhomogeneous turbulent fields with prescribed autocorrelation functions, Commun.\ Numer.\ Meth.\ Engng. 23~(1) (2007) 35--43.

\bibitem{Mercier:2015}
R.~Mercier, T.~Schmitt, D.~Veynante, B.~Fiorina, The influence of combustion sgs submodels on the resolved flame propagation.{A}pplication to the les of the cambridge stratified flames, Proc.\ Comb.\ Inst. 35~(2) (2015) 1259--1267.

\end{thebibliography}

% -------------------------------------------------------------------- %
% -------------------------------------------------------------------- %
% -------------------------------------------------------------------- %

\newpage

\small
\baselineskip 10pt

% -------------------------------------------------------------------- %
% -------------------------------------------------------------------- %
% -------------------------------------------------------------------- %

% -------------------------------------------------------------------- %
% -------------------------------------------------------------------- %
% -------------------------------------------------------------------- %

\end{document}